\documentclass{article}

\usepackage{arxiv}

\usepackage[utf8]{inputenc} 
\usepackage[T1]{fontenc}    
\usepackage{hyperref}       
\usepackage{url}            
\usepackage{booktabs}       
\usepackage{amsfonts}       
\usepackage{nicefrac}       
\usepackage{microtype}      
\usepackage{lipsum}
\usepackage{graphicx}
\usepackage{hyperref}
\usepackage{natbib}
\usepackage{times}
\usepackage{url}

\title{Algorithmic Fairness in AI Surrogates for End-of-Life Decision-Making}

\author{
 Muhammad Aurangzeb Ahmad \\
The Department of Computing \& Software Systems \\
University of Washington Bothell\\
Harborview Medical Center, UW Medicine\\
  \texttt{maahmad@uw.edu} \\ \\
}
\begin{document}
\maketitle

\begin{abstract}
Artificial intelligence surrogates are systems designed to infer preferences when individuals lose decision-making capacity. Fairness in such systems is a domain that has been insufficiently explored. Traditional algorithmic fairness frameworks are insufficient for contexts where decisions are relational, existential, and culturally diverse. This paper explores an ethical framework for algorithmic fairness in AI surrogates by mapping major fairness notions onto potential real-world end-of-life scenarios. It then examines fairness across moral traditions. The authors argue that fairness in this domain extends beyond parity of outcomes to encompass moral representation, fidelity to the patient’s values, relationships, and worldview.
\end{abstract}


\section{Introduction}
The increased application of artificial intelligence in healthcare is transforming not only diagnostic and predictive medicine but also the moral landscape of clinical decision-making. Among the most ethically charged applications is the use of AI surrogates, systems trained on a patient’s medical, behavioral, and psychosocial data to make or recommend medical decisions when the patient is incapacitated. These systems aspire to simulate what a patient would have wanted, thereby extending the principle of substituted judgment into the digital domain. Although the use of AI IN end-of-life decision making \cite{ahmad2018death} and surrogate systems has been the subject of growing scholarly debate, such systems have not yet been deployed in clinical practice owing to their ethical and technical limitations. Among these, one critical yet underexplored concern is the question of algorithmic fairness. This paper seeks to address this omission by examining how fairness frameworks intersect with the moral, cultural, and procedural dimensions of end-of-life decision-making.

In traditional medical ethics, surrogate decision-making is an act of moral interpretation. A human surrogate, a spouse, relative, or legally appointed agent, is expected to represent the patient’s values and prior wishes to the best of their knowledge \citep{beauchamp2013principles, sudore2010preferences}. This process, while imperfect, is relational in nature. It emerges from empathy, memory, and shared life experience. In contrast, an AI surrogate could rely on patterns learned from data, clinical variables, demographic information, linguistic markers in clinical notes, and possibly the patient’s digital footprint. This substitution of predictive reasoning for moral reasoning introduces a profound epistemic and ethical tension: Can a machine truly represent the will of a person, or does it merely reproduce the statistical regularities of populations?

The promise of AI surrogates lies in their potential to mitigate known problems in human surrogate decision-making. Studies have shown that human surrogates often misjudge patient preferences, projecting their own hopes, fears, or cultural expectations \citep{shalowitz2006accuracy, fagerlin2001}. AI systems, proponents argue, could rely on a broader data base of patient behavior, prior medical choices, or documented preferences to provide more consistent and less emotionally biased recommendations \citep{bennett2021, grote2022}. However, as numerous scholars have cautioned, consistency without fairness is merely automation of inequity \citep{obermeyer2019, benjamin2019race}. The moral legitimacy of an AI surrogate should depend not only on its predictive accuracy but also on the promise of its fairness across individuals and groups, its transparency of reasoning, and its attunement to cultural diversity in conceptions of the good death.

This current paper seeks to articulate and analyze the ethical, cultural, and epistemic dimensions of fairness in AI surrogates for end-of-life care especially in the context of Do-Not-Resuscitate (DNR) decision-makinge. We argue that conventional algorithmic fairness frameworks, designed for distributive contexts are inadequate for settings where decisions are existential and relational. End-of-life fairness must be understood not only as equitable distribution of outcomes but as moral adequacy of representation, the degree to which an AI system honors the patient’s values, relationships, and cultural worldview.  We map the major notions of algorithmic fairness to specific ethical risks in AI-assisted DNR decisions, ranging from bias propagation and unequal recommendation rates to loss of cultural meaning and interpretability. Normatively, we argue for a moral pluralism framework of fairness, one that acknowledges multiple valid moral vocabularies and embeds cultural context into data governance, model interpretation, and clinical deployment. The paper concludes by offering design and governance implications for developing culturally attuned AI surrogates that respect both algorithmic integrity and human dignity.

\section{AI Surrogates and End-of-Life Decision-Making}
Surrogate decision-making is a cornerstone of clinical ethics when patients cannot express their preferences. The traditional ethical standards are \textit{substituted judgment} i.e., where surrogates attempt to make the decision the patient would have made and the \textit{best-interest standard}, which aims to maximize patient welfare when preferences are unknown \citep{beauchamp2013principles}. However, empirical studies show that surrogates often misrepresent patient wishes. Accuracy rates rarely exceed 68\%, even among close relatives \citep{shalowitz2006accuracy}. Emotional stress, cognitive bias, and limited communication exacerbate these errors \citep{fagerlin2001, sudore2010preferences}. Additionally, cultural norms profoundly shape how autonomy and authority are interpreted. In some societies, family members are expected to decide collectively, while in others, autonomy implies individual choice.

AI surrogates aim to assist or partially automate this moral translation process by drawing on large-scale data sources, including electronic health records, prior treatment choices, natural-language notes, wearable data, and recorded conversations \citep{bennett2021, grote2022}. Through statistical modeling and machine learning, such systems estimate what a patient with similar characteristics, beliefs, or prior behaviors would prefer under comparable conditions. This predictive capacity could, in theory, reduce emotional burden and human inconsistency. However, it also shifts the epistemic foundation of decision-making from relational empathy to probabilistic inference. The substitution of data-driven inference for personal knowledge introduces several risks. First, data are not neutral. Medical records capture institutional perspectives, and they often underdocument the preferences of marginalized patients \citep{obermeyer2019}. Second, a model trained on population-level data may systematically underrepresent minority value systems, resulting in \textit{epistemic injustice}, the misrepresentation or silencing of certain moral voices \citep{benjamin2019race}. That said, individual models based on a particular patient's data would fare much better. Third, prediction without context may lead to moral overreach, where algorithmic confidence is mistaken for ethical authority.

Among EOL choices, Do-Not-Resuscitate (DNR) orders exemplify the moral and procedural complexity of surrogate decision-making. Empirical research shows that DNR orders are unevenly distributed across racial, socioeconomic, and geographic lines \citep{barnato2009, morrison2000}. Black and Hispanic patients are statistically less likely to have DNR orders compared with white patients, reflecting both historical mistrust and cultural differences in end-of-life preferences. If AI systems are trained on such imbalanced data, they may reproduce or amplify these disparities. Because DNR errors are morally asymmetric, falsely recommending a DNR can irreversibly violate a patient’s will, standard fairness metrics that treat false positives and false negatives symmetrically are ethically inadequate. The case of AI-assisted DNR decision-making thus foregrounds the need for a broader conception of fairness. Technical definitions of fairness focus on outcome parity or equal error rates, yet in EOL care, fairness entails moral representation: faithfully expressing the patient’s worldview and values. The central question becomes not only, ``Is the model unbiased?'' but ``Whose moral universe does the model inhabit?''

\section{Algorithmic Fairness in End-of-Life}

Artificial intelligence systems used for clinical decision support increasingly rely on different fairness frameworks drawn from computer science, philosophy, and law.  These fairness paradigms have different trade-offs and are often incompatible with each other. Statistical parity may conflict with individual or causal fairness, equalized odds may ignore relational asymmetries. In the context of AI surrogates, no single notion of fairness can capture moral adequacy. This section outlines major fairness paradigms and evaluates their applicability to AI surrogates.

\subsection{Demographic parity and statistical fairness}
Demographic parity, also called statistical parity, requires that an algorithm’s output be independent of protected attributes such as race or gender. In practice, this means that the probability of a DNR recommendation should be similar across groups. Although intuitive, this criterion can mask important contextual differences. Group-level parity does not necessarily indicate fairness if underlying data reflect historical or cultural variations in preferences. For instance, demographic differences in advance care planning may arise from mistrust or community norms rather than bias in model processing \citep{barnato2009}. Achieving parity without understanding causal context risks moral homogenization \citep{binns2018, green2018}. Consider a hospital dataset in which Black patients receive DNR orders at lower rates than white patients \citep{barnato2009}. An algorithm trained on such data might learn that “being Black” predicts a lower likelihood of DNR preference. Enforcing demographic parity could equalize DNR recommendation rates across races. However, doing so without understanding the roots of disparity, historical mistrust, communication gaps, or cultural values, risks introducing new ethical errors. Statistical parity without causal insight can amount to moral homogenization \citep{green2018}. Fairness requires contextual audits that distinguish bias from authentic heterogeneity.

\subsection{Equal opportunity and equalized odds}
Equal opportunity and equalized odds extend the notion of fairness beyond outcomes to encompass error rates. Under the principle of equal opportunity, individuals who would benefit from a given outcome, such as the alignment between an AI system’s recommendation and the patient’s actual preferences, should have an equal likelihood of receiving that outcome across all groups. Equalized odds adds the requirement that false positive rates be equal as well \citep{hardt2016}. However, EOL decisions are asymmetrical in moral weight, a false positive DNR (withholding life-sustaining treatment) carries more severe ethical cost than a false negative. This asymmetry challenges the symmetry assumptions of traditional metrics \citep{green2018}. Fairness here should reflect non-maleficence rather than balance alone. Suppose an AI model predicts a high likelihood of DNR preference for an elderly patient. If the system’s false positive rate is disproportionately high for certain groups, the consequences differ morally. A false positive DNR can irreversibly deny life-sustaining treatment, whereas a false negative can still be reviewed by clinicians or families. Standard equal opportunity frameworks that balance error rates may therefore be ethically insufficient \citep{hardt2016}. Weighted loss functions that penalize false positives more severely operationalize fairness as non-maleficence rather than mere parity.

\subsection{Individual fairness}
Individual fairness posits that similar individuals should be treated similarly \citep{dwork2012}. In clinical contexts, defining similarity is ethically fraught. Two patients may share identical clinical features but diverge in moral reasoning or spiritual beliefs. Embedding such value-based distinctions into distance metrics risks overfitting morality into data space. Individual fairness in AI surrogates therefore requires not only technical similarity but moral contextualization, acknowledging heterogeneity in how people derive meaning from illness. Consider the following: Two patients of similar clinical profiles may differ in moral reasoning, one guided by autonomy, another by family or religious duty. Treating them “similarly” in algorithmic terms would constitute moral erasure. Individual fairness requires incorporating value-sensitive features, such as recorded spiritual preferences or statements about comfort, without violating privacy.  AI surrogates would thus need to recognize the individuality of moral reasoning, not just clinical similarity.

\subsection{Counterfactual fairness}
Counterfactual fairness ensures that an algorithm’s output would not change if a protected attribute were altered in a counterfactual world \citep{kusner2017}. Applied to DNR decisions, this would require that changing a patient’s race while holding all else constant does not affect the model’s recommendation. However, identity cannot be neatly abstracted from lived experience. A more appropriate approach is \textit{causal path-specific fairness}, which distinguishes legitimate causal pathways (e.g., cultural values shaping end-of-life preferences) from illegitimate ones (e.g., systemic documentation bias) \citep{chiappa2022}.

\subsection{Procedural fairness}
Procedural fairness emphasizes the integrity and transparency of the decision-making process rather than the fairness of outcomes alone. In the context of AI surrogacy, this includes transparency, explainability, interpretability, and mechanisms for recourse. Stakeholders, clinicians, patients, and families, must be able to understand how recommendations are generated, what data sources are used, and what uncertainties or value judgments are embedded in the model \citep{selbst2019, ahmad2018interpretable}. Explainability in this domain extends beyond technical transparency. It is not sufficient for an AI surrogate to provide numerical justifications or feature weights, explanations must be comprehensible to non-technical users and sensitive to the emotional gravity of end-of-life decisions. Families should be able to grasp, in accessible language, why a system has inferred that a DNR recommendation aligns with the patient’s prior values or documented choices. Similarly, clinicians should be able to interrogate which features, such as prior refusals of ventilation, language in clinical notes, or recorded preferences, most strongly influenced the recommendation. This kind of layered explainability allows both epistemic accountability and ethical reflection.

Additionally, procedural fairness demands that stakeholders have the ability to contest, appeal, or override algorithmic outputs. AI surrogates must include clearly defined recourse mechanisms that allow clinicians and family members to challenge recommendations when they conflict with contextual judgment or new information. Appeals should trigger structured human deliberation, preferably involving ethics committees or oversight boards, rather than unaccountable technical exceptions \citep{grote2022}. Lastly, procedural fairness anchors accountability in the sociotechnical system surrounding the algorithm rather than in the algorithm alone. It requires continuous documentation, transparent governance of model updates, and independent auditing.

\subsection{Relational and recognition fairness}
Traditional fairness metrics often overlook the importance of recognition and respect. Drawing from political philosophy, relational fairness focuses on how individuals are regarded within systems of power \citep{binns2018, fraser1997, honneth1995}. In EOL contexts, relational fairness requires acknowledging patients as persons embedded in cultural and familial relationships. Failure to recognize moral worldviews, such as reading silence as consent or undervaluing collective deliberation, constitutes a form of misrecognition and moral injury. Designing for relational fairness entails ensuring that AI surrogates communicate and justify their recommendations in ways consistent with patients’ moral grammars.

\subsection{Temporal fairness}
EOL preferences are not static. Patients may revise their wishes as illness progresses, pain intensifies, or family circumstances change. Temporal fairness therefore refers to the preservation of alignment between model assumptions and evolving patient values over time \citep{heidari2019, kaye2015}. Systems should support dynamic consent and time-stamped value updates. Temporal fairness also applies at the institutional level, requiring periodic recalibration as cultural attitudes toward death and technology shift. A patient may initially choose aggressive treatment but later prioritize comfort care as illness progresses. Temporal fairness ensures that models account for such value drift by incorporating time-stamped consent updates and longitudinal learning \citep{heidari2019, kaye2015}. Without temporal tracking, AI surrogates risk committing the "frozen self" error, representing outdated values as current preferences. Clinically, this means fairness is achieved not through static prediction but through continuous moral recalibration.

\subsection{Intersectional fairness}
Intersectional fairness accounts for overlapping systems of disadvantage \citep{crenshaw1991}. In healthcare, algorithmic bias often compounds across identities: for example, older Black women may face both gendered and racialized disparities in documentation and care. Fair AI surrogates must audit not only marginal distributions but also intersectional subgroups, ensuring representational adequacy across demographic and cultural axes \citep{buolamwini2018}. Intersectional auditing should inform data collection, model evaluation, and governance. Epistemic fairness extends beyond bias correction to question whose moral frameworks define the "ground truth." Western datasets often encode individualistic notions of autonomy, while Confucian or Islamic traditions emphasize family and divine relationality. To achieve epistemic fairness, models must be locally contextualized: through community engagement, participatory co-design, and ethical localization protocols that adapt interpretive norms to local moral ecologies \citep{mohamed2020, carroll2020}.

\subsection{Epistemic and ontological fairness}
Epistemic fairness asks whose knowledge and worldview shape the algorithm’s design and data ontology \citep{crawford2021, mohamed2020}. In healthcare datasets, Western biomedical categories often dominate, marginalizing non-Western moral frameworks. Ontological fairness extends this concern to representation: whether categories such as “patient autonomy” or “quality of life” carry the same meaning across cultures. Ensuring epistemic fairness requires participatory design, transparency in value choices, and the inclusion of culturally diverse experts in data annotation and model interpretation.

\section{Cross-Cultural Perspectives on Fairness in AI Surrogacy}

Algorithmic fairness, as developed in Western context, often presumes a liberal individualist moral ontology i.e., that persons are autonomous agents, fairness is equality of treatment, and justice arises from impartial procedure. While useful, this framing is culturally contingent. End-of-life decisions reveal divergent moral grammars across societies. This section explores how major ethical traditions conceptualize fairness, autonomy, and relationality, and how these perspectives can inform AI surrogate design.

Western bioethics, grounded in the principles of autonomy, beneficence, non-maleficence, and justice \citep{beauchamp2013principles}, conceptualizes fairness primarily as procedural impartiality and respect for individual rights. The ideal surrogate acts as a neutral proxy who enacts the patient’s self-determined choices. This aligns with algorithmic fairness paradigms like equal opportunity or individual fairness, which emphasize consistent treatment across cases. However, critics note that such frameworks neglect relational and affective dimensions of care. Care ethics and narrative medicine counter this by situating fairness in empathy and storytelling, emphasizing understanding the patient’s lived narrative rather than merely honoring abstract autonomy \citep{charon2006}.

In Confucian moral philosophy, fairness is relational harmony rather than impartial equality. Ethical action arises from fulfilling one’s roles, child, parent, physician, in accordance with \textit{ren} (humaneness) and \textit{li} (ritual propriety) \citep{tu1985}. End-of-life decisions are typically collective, guided by family consultation rather than individual autonomy \citep{ho2008}. For AI surrogates, this might imply designing systems that consider the moral weight of family relationships and context-sensitive reasoning. Fairness entails preserving relational balance and minimizing disharmony. Rather than predicting a singular “preference,” the AI surrogate might facilitate negotiation among family members, respecting hierarchy and reciprocity.

Islamic ethics grounds fairness in divine justice (\textit{adl}) and mercy (\textit{rahmah}), framing human life as a trust (\textit{amana}) from God rather than personal possession \citep{aksoy2001, sachedina2009}. Decisions about resuscitation or life support must balance human stewardship with acceptance of divine decree. The concept of \textit{shura} (consultation) provides a deliberative model: ethical decisions should emerge from mutual consultation among family, physicians, and patient's own preference in light of the extended Islamic framework. AI surrogates operating in Muslim contexts should thus encode such values.

In Hindu and Buddhist traditions, fairness is aligned with \textit{dharma} (righteous duty) and \textit{karuṇā} (compassion) \citep{keown2005, bhawuk2020}. Moral reasoning is situational, balancing roles and intentions within karmic continuity. Fairness is achieved through non-harm (\textit{ahimsa}) and equanimity. For AI surrogates, this would suggest an ethics of proportionality: the system should guide decisions that minimize suffering and attachment while honoring duty.  Indigenous ethics often center reciprocity, interdependence, and respect for the continuity between human and ecological communities \citep{deloria2006, carroll2020}. Fairness entails honoring communal sovereignty over data and decision-making. The CARE principles, Collective benefit, Authority to Control, Responsibility, and Ethics, extend the FAIR data standards by asserting Indigenous data rights \citep{carroll2020}. For AI surrogates, this means that datasets derived from Indigenous populations should be governed by those communities, and that model behavior should align with their cosmologies of balance and stewardship. Decolonial perspectives caution against moral extractivism, the appropriation of ethical insights without structural reciprocity \citep{mohamed2020}.

Despite differences, these traditions converge on three shared insights. First, fairness is relational: it arises from the right configuration of relationships, not only from equal treatment. Second, fairness is dialogical: it emerges through consultation, ritual, or story, not algorithmic determination. Third, fairness is situated: moral reasoning depends on context, history, and spiritual ontology. Therefore, AI surrogates must be designed as moral mediators that adapt to diverse ethical vocabularies rather than enforcing universalist norms. This pluralistic orientation supports UNESCO’s call for “culturally responsive AI ethics” \citep{unesco2021}.
\section{Design Implications and Governance}
In the previous sections we discussed that fairness cannot be achieved solely through model optimization. It must be embedded in the sociotechnical systems that produce, deploy, and govern AI. This section outlines the design implications, governance structures, and accountability mechanisms necessary to operationalize fairness as moral attunement. Fairness is not a property of algorithms but an emergent quality of the system in which they operate. Governance therefore must span three interdependent domains, technical architecture, clinical workflow, and cultural-ethical oversight. Achieving fairness requires what might be termed \textit{ethical infrastructure engineering}, building organizational mechanisms that support continual moral reflection and adaptation. Fairness must be maintained throughout the AI lifecycle, from data collection to post-deployment auditing. Each phase carries distinct ethical risks and governance tasks, as summarized in Table \ref{tab:ai_cycle}:

\begin{table}[htbp]
\centering
\caption{AI Model Building Ethical Risks and Governance}
\label{tab:ai_cycle}
\begin{tabular}{|p{3cm}|p{6cm}|p{6cm}|}
\hline
\textbf{Phase} & \textbf{Ethical Risks} & \textbf{Governance Tasks} \\
\hline
Data acquisition & Representation bias, privacy violation & Consent design, provenance tracking, community oversight \\
\hline
Model development & Feature selection bias, moral reductionism & Participatory co-design, value-sensitive modeling \\
\hline
Deployment & Contextual mismatch, opacity & Human-in-the-loop integration, explainability interfaces \\
\hline
Post-deployment & Model drift, cultural insensitivity & Continuous auditing, recourse mechanisms, cultural recalibration \\
\hline
\end{tabular}
\end{table}

Data are morally charged artifacts. Variables such as “non-compliance,” “comfort care,” or “chaplain visits” encode institutional values and social hierarchies \citep{crawford2021}. Ethical fairness requirements would  include \textit{moral metadata}, documenting who collected the data, under what circumstances, and whose values were expressed. For example, a DNR note should record who initiated the discussion, the language used, and the presence of interpreters or family. This transparency enables later audits for cultural misrepresentation.

Technical data governance frameworks such as the FAIR principles (Findable, Accessible, Interoperable, Reusable) should be supplemented by the CARE principles (Collective benefit, Authority to control, Responsibility, Ethics) \citep{carroll2020}. This integration ensures that data management respects not only technical interoperability but also cultural sovereignty. For communities historically marginalized in healthcare data, CARE alignment protects against extraction and promotes shared benefit.

Patient values evolve with illness trajectories. Consent should therefore be treated as a dynamic process rather than a one-time event \citep{kaye2015}. Interfaces enabling patients or their surrogates to periodically review, update, or withdraw their data ensure \textit{temporal fairness}. This approach prevents the “frozen self” problem, using outdated data to represent current values, and upholds patient autonomy over time.

Developers must distinguish between clinical, contextual, and moral features. Clinical features describe physiology, contextual features reflect social determinants, and moral features capture value preferences. Each category warrants different levels of consent and ethical review. Loss functions should be adjusted to reflect moral asymmetry, penalizing false-positive DNR recommendations more heavily than false negatives \citep{green2018}. In addition, each model should include a \textit{Moral Assumption Statement}, a short documentation note describing the ethical trade-offs embedded in its design.

AI surrogates must operate as decision aids, not decision makers. Every recommendation should pass through clinicians and ethics oversight \citep{grote2022}. Formal recourse mechanisms should be established, where contested outputs trigger ethics review. Each decision should leave an auditable trail documenting how algorithmic insights were accepted, modified, or rejected. This traceability sustains procedural fairness and legal accountability.

Explainability in EOL contexts should not be reduced to technical transparency. It must become an ethical dialogue. For example, an AI surrogate might communicate, “Based on prior notes indicating a preference for comfort and avoidance of ventilation, this recommendation reflects a 90\% probability that comfort care aligns with the patient’s wishes. Would you like to review the supporting evidence?” Such dialogic explanations foster trust, empathy, and collaborative deliberation.

Post-deployment governance requires routine fairness audits that combine quantitative and qualitative metrics. These may include:

\begin{itemize}
\item Demographic outcome parity gaps (less than 5\%)
\item Group-specific false positive and false negative rates
\item Temporal drift in model calibration and value alignment
\item Stake-holder and clinician trust ratings
\item Representation coverage across intersectional subgroups
\end{itemize}

Fairness audits should culminate in \textit{Fairness Impact Assessments}, publicly available reports akin to environmental impact statements \citep{holstein2019}. This institutionalizes transparency and social accountability.

AI surrogates fall under the category of high-risk decision systems. Hospitals must establish clear chains of responsibility identifying who authorized, reviewed, and implemented algorithmic recommendations. Regulators should align with global frameworks such as the EU AI Act and UNESCO’s AI Ethics Recommendation \citep{unesco2021}. Internal oversight can integrate existing structures, Institutional Review Boards, Clinical Ethics Committees, and Data Protection Officers, to create a unified ethics governance ecosystem. At the system level, fairness can be embedded through modular design. \textit{Ethics middleware} can act as a configurable layer between the predictive engine and the user interface, enforcing context-specific fairness parameters. Version control logs should track ethical as well as technical changes, with approval required before deployment. 
\section{Discussion and Future Directions}
The application of fairness frameworks to AI surrogates highlights the limits of conventional algorithmic definitions. Fairness cannot be reduced to parity in outcomes or balance in error rates. In end-of-life contexts, fairness functions as a form of \textit{moral attunement}, the capacity of a sociotechnical system to resonate with human values, vulnerabilities, and cultural plurality. Whereas fairness metrics quantify equality, moral attunement recognizes empathy, humility, and relational context as integral to just decision-making. The challenge is to design AI surrogates that can participate in moral reasoning without overstepping human judgment.

It is also possible that next-generation models will integrate voice, facial expression, social media, and longitudinal health data to infer patient intent with greater fidelity. While this promises more contextually aware recommendations, it also risks creating a form of “predictive personhood,” where the surrogate embodies an algorithmic reconstruction of the patient. The moral challenge will be ensuring that such reconstructions remain interpretable and faithful without crossing into simulation or posthumous autonomy. Conversational AI surrogates: Within the next decade, families may interact with empathetic conversational agents that explain treatment options using the patient’s own linguistic style, derived from past writings or speech data. While this may provide comfort and familiarity, it also blurs the boundary between assistance and emotional manipulation. Ethical governance will need to address the authenticity of voice, the right to digital silence, and the prevention of persuasive bias. It may turn out that the use of such systems would be unwarranted given the high stakes involved. 

These technological evolutions will generate new moral tensions e.g., As AI surrogates become capable of reproducing speech and reasoning patterns of patients, questions will arise about continuity of personhood. Can a digital surrogate make morally binding statements after a person’s death? How should we regulate consent for posthumous digital decision-making? There is also the question of ownership of the data e.g., if a patient’s preferences and moral narratives are encoded into a model, who owns that representation—the patient, the family, or the healthcare institution? Data governance frameworks must expand to include ownership of moral and relational data, not just clinical data. Additionally, emotionally intelligent surrogates could unintentionally manipulate decision-making by using empathetic tone or persuasive framing. This may be yet another reason to not have these system take on a human persona. Fairness here involves affective neutrality and emotional transparency—ensuring that the system’s tone supports deliberation rather than coercion.

Closer to the present,  fairness metrics translate moral principles into mathematical form, but they cannot fully capture the existential and affective dimensions of dying. The moral meaning of a DNR recommendation depends on histories of care, family relationships, and spiritual interpretations of death. An AI surrogate may estimate probabilities of preference but cannot feel obligation or remorse. The goal is not automation of ethics but augmentation of moral deliberation, designing systems that foster reflection, not replacement. EOL decision-making involves deep uncertainty not only about outcomes but about values themselves. Patients often clarify their moral priorities only through the lived process of illness. Consequently, fairness requires flexibility to accommodate evolving self-understanding. AI surrogates should therefore function as reflective companions that help patients and families explore values before crises occur. Narrative-driven data collection, where patients articulate stories about meaning and legacy, can enhance the moral fidelity of predictions. Fairness, in this sense, supports epistemic growth: helping individuals come to know what matters most to them.

If AI surrogates are introduced across cultural contexts, fairness must be reinterpreted through local moral grammars. A universal fairness metric risks imposing Western individualism as moral default \citep{mohamed2020}. Instead, ethical localization should guide deployment. This process adapts system parameters, explanation styles, and consent protocols to reflect community values. Achieving fairness thus requires \textit{moral interoperability}, the ability of AI systems to translate between ethical frameworks without erasing their differences. Additionally, empirical studies are needed to evaluate how fairness frameworks perform in real clinical contexts for DNR related end of life scenarios. Future research should:
\begin{itemize}
\item Quantify demographic and intersectional disparities in AI-assisted DNR recommendations;
\item Assess family and clinician perceptions of fairness and trust;
\item Compare outcomes between culturally localized and generic models;
\item Investigate how dynamic consent interfaces affect patient satisfaction and alignment.
\end{itemize}
Mixed-methods designs combining statistical audits with ethnographic observation will capture both quantitative bias and qualitative moral resonance. Simulation environments, ethical sandboxes, can safely explore fairness trade-offs before real-world deployment.

Governments and health institutions should recognize AI surrogates as high-stakes moral technologies requiring ethical regulation. Policy should mandate fairness and cultural impact assessments analogous to clinical trials. Regulatory bodies can establish transparency requirements, human oversight clauses, and mandatory reporting of fairness audits. Hospitals should institutionalize Ethics Translation Boards to oversee design, deployment, and monitoring. Funding agencies can incentivize fairness by requiring participatory ethics plans in AI research proposals. Such multi-level governance embeds fairness not as compliance but as moral infrastructure.

The use of AI to represent incapacitated persons raises important philosophical questions about personhood. When an algorithm speaks for someone who can no longer speak, it becomes a moral surrogate as well as a computational one. Fairness here means fidelity to the person’s moral identity, not only statistical accuracy. This reframes AI as a medium of representation, of memory, voice, and moral agency. Ensuring fairness in this domain thus protects the integrity of human selfhood against reduction to data abstraction. Ultimately, fairness in AI surrogates must be understood as a form of care. It can be thought of as a sustained commitment to relational justice and epistemic humility. Fairness is achieved not by eliminating difference but by listening across it. The moral success of AI surrogates will not be measured only by calibration curves but by whether they help clinicians and families face death with greater clarity, compassion, and dignity. 
\section{Conclusiony}

The investigation of algorithmic fairness in AI surrogates for end-of-life decision-making brings to the forefront a central question of biomedical ethics, what does it mean to act justly when another person’s dignity and story are at stake. This paper has argued that fairness, in such morally charged contexts, is not a computational property but a moral grammar, a language through which societies articulate the worth of persons at the threshold of death. In many domains of machine learning, fairness is treated as an optimization constraint among others. Yet in surrogate decision-making, fairness becomes the condition of moral legitimacy. An AI surrogate that recommends a DNR order or continued life support effectively speaks on behalf of the patient. Fairness, therefore, entails fidelity to that person’s values, relationships, and worldview. It is a form of moral representation, not just a statistical goal. Metrics alone cannot capture whether the AI has respected autonomy, mercy, or family duty. 

Cultural analysis of this problem reveals that fairness is interpreted differently across moral traditions. Western liberalism emphasizes individual rights and procedural impartiality, while Confucian ethics centers on relational harmony, Islamic bioethics on divine trusteeship, Ubuntu on communal solidarity, and Dharmic ethics on compassionate balance. Indigenous and decolonial perspectives highlight collective sovereignty and reciprocity with nature. Despite diversity, all converge on a relational understanding of fairness, grounded in care, humility, and dialogue. Fair AI surrogates must therefore act as moral mediators capable of adapting to plural moral vocabularies rather than enforcing a single universal code.

Operationalizing fairness requires institutional design. Ethics Translation Boards, dynamic consent interfaces, CARE-aligned data stewardship, and continuous fairness audits turn moral aspiration into governance practice. These structures ensure that fairness is not a static compliance goal but a living, adaptive process. Hospitals must treat ethical oversight as infrastructure, embedding reflection and accountability at every stage of data collection, model training, deployment, and review. When fairness becomes institutional habit, it transforms from principle into practice.
 No algorithm can replace human judgment in moral matters. AI surrogates can estimate likelihoods, but they cannot bear moral responsibility. The goal, therefore, is co-responsibility: machines augmenting human deliberation, not substituting for it. Fairness means preserving the space for human empathy and uncertainty, ensuring that technology amplifies rather than silences moral agency. The fairest AI surrogate is one that invites conversation, admits doubt, and leaves room for care.

Future work should pursue three directions, empirical validation of fairness frameworks in clinical trials, simulation-based evaluation of moral trade-offs, and interdisciplinary education combining AI design with cross-cultural bioethics. Such efforts will move fairness research beyond technical compliance toward epistemic inclusivity and social trust. In the final analysis, fairness in AI surrogates is best understood as a form of care. To be fair is to attend to the vulnerable, to listen across difference, and to honor human dignity in the presence of uncertainty. A fair system does not simply balance datasets, it sustains moral relationships. If future AI surrogates can help families and clinicians navigate death with greater compassion and clarity, they will have achieved a deeper fairness, one that unites reason with empathy, and technology with humanity.

It is important to clarify that our discussion of AI surrogates does not imply advocacy for their deployment as an inevitability. Rather, this work is an exploration of their conceptual, ethical, and technical possibilities as far as fairness is concerened. The aim is to anticipate the moral questions such systems would raise should they ever be developed, not to endorse their use in clinical practice. If empirical evidence and rigorous ethical evaluation ultimately suggest that such systems are unsafe, untrustworthy, or incompatible with human dignity, then the appropriate conclusion would be that they should not be used. Ethical inquiry, in this sense, includes the possibility of refusal.

\bibliographystyle{apalike}
\bibliography{main}

\end{document}